\documentclass[universe,article,accept,moreauthors,pdftex]{mdpi}
\firstpage{1}

\pubvolume{xx} \issuenum{1} \articlenumber{5} \pubyear{2020} \copyrightyear{2020}
\history{Received: 13 July 2020; Accepted: 5 August 2020; Published: date}
\setitemize{parsep=6pt,itemsep=0pt,leftmargin=*,labelsep=5.5mm}
\setenumerate{parsep=6pt,itemsep=0pt,leftmargin=*,labelsep=5.5mm}
\setlist[description]{itemsep=0mm}

\usepackage{soul,upgreek}
\usepackage{booktabs}
\usepackage{multirow}
\usepackage{microtype}
\usepackage{amssymb}                 % Extra maths symbols
\def\be{\begin{equation}} \def\ee{\end{equation}}
\def\bal#1\eal{\begin{align}#1\end{align}}
\def\ra{\rightarrow}

\def\de{\Delta}
\def\ms{\,M_\odot}
\def\mmax{M_\text{max}}
\def\fm3{\,\text{fm}^{-3}}
\def\gc3{\,\text{g/cm}^3}
\def\rdu{\rho_\text{DU}}
\def\mdu{M_\text{DU}}
\def\r1s0{\rho_{1S0}}
\def\m1s0{M_{1S0}}

\Title{Nuclear Pairing Gaps and Neutron Star Cooling}

\Author{%J.-B. Wei, G. F. Burgio and H.-J. Schulze,
Jin-Biao Wei $^*$,     %$^{1,\dagger,\ddagger}$\orcidA{},
Fiorella Burgio and         %$^{1,\ddagger}$ and
Hans-Josef Schulze          %$^{2,}$*
}
\AuthorNames{J.-B. Wei, G. F. Burgio and H.-J. Schulze}

\address[1]{% Add [1] after \address if there is only one affiliation
INFN Sezione di Catania,
Dipartimento di Fisica e Astronomia, Universit\`a di Catania,
Via Santa Sofia 64, \mbox{95123 Catania}, Italy;
burgio@ct.infn.it (F.B.);
schulze@ct.infn.it (H.-J.S.)
}

\corres{\hangafter=1 \hangindent=1.05em \hspace{-0.82em}
Correspondence: jinbiao.wei@ct.infn.it}
\abstract{
We study the cooling of isolated neutron stars
with particular regard to the importance of nuclear pairing gaps.
A microscopic nuclear equation of state derived in the
Brueckner-Hartree-Fock approach is used
together with compatible neutron and proton pairing gaps.
We then study the effect of modifying the gaps
on the final deduced neutron star mass distributions.
We find that a consistent description of all current cooling data can be achieved
and a reasonable neutron star mass distribution can be predicted
employing the (slightly reduced by about 40\%)
proton 1S0 Bardeen-Cooper-Schrieffer (BCS) gaps
and no neutron 3P2 pairing.
}


\begin{document}

\section{Introduction}

A very important effect of nuclear superfluidity in a neutron star (NS)
is the suppression of standard neutrino cooling processes
and the appearance of new ones,
which then compete with each other~\cite{2001rep,yako14,bryn18,armen19}.
The superfluidity has therefore a decisive influence
on the temperature evolution of an isolated NS,
and this can be compared with known observational data
in order to deduce constraints on the various pairing gaps
and also on the nuclear equation of state (EOS).
This article is dedicated to this problem,
and we will perform a detailed analysis of NS cooling with the above~goal.

It is currently still not clear whether all NSs have a purely
nucleonic substructure, that is,
can be considered to be built of individual neutrons and protons,
or whether heavy NSs hide exotic baryons like hyperons
or even deconfined quark matter in their extremely dense interior.
In this work we follow the simple first assumption and consider
purely nucleonic NSs.
We model their internal structure by a theoretical EOS that has been
derived within the Brueckner-Hartree-Fock (BHF) many-body method,
fulfilling all current constraints imposed by observational data
from nuclear structure, heavy-ion collisions, NS global properties,
and recently NS merger events~\cite{merger,merger2,merger3}.
Within this framework we then investigate the cooling evolution of NSs,
and in particular the effect of the proton 1S0
and the neutron 3P2 pairing~gaps.

This investigation has been carried out in several previous
publications~\cite{2016MNRAS,2018MNRAS,2019MNRAS,2020MNRAS},
and we refine here our analysis regarding the constraints on the
pairing gaps deduced from comparison with cooling data.
In fact we have previously concluded that a good reproduction
of all current cooling data is possible by assuming the
Bardeen-Cooper-Schrieffer (BCS) gap in the proton 1S0 channel,
but not allowing pairing in the neutron 3P2 channel.
However, the~BCS approximation disregards any medium effects on the gaps,
which are not supposed to be small in the dense NS matter.
An accurate quantitative theoretical computation of such effects
is however still very difficult or impossible,
and therefore we investigate in this work empirically the effects
of such modifications on the cooling evolution
in order to identify possible constraints that may be obtained in this way.
In particular, we concentrate in this article on the NS mass distribution
that can be obtained by comparing theoretical cooling curves
with the currently available set of cooling data~\cite{Beznogov1}
(We do not yet utilize the very recent update of these data~\cite{Potek20}
in this work),
and which is therefore a functional of the pairing~gaps.

This paper is organized as follows.
In Section~\ref{s:eos} we give a brief overview of the theoretical framework,
namely the BHF formalism adopted for the nuclear EOS,
the various cooling processes,
and the related nucleonic pairing gaps.
Section~\ref{s:res} is devoted to the presentation and discussion
of the results for stellar structure,
the cooling diagrams, and~the derived mass distribution.
Conclusions are drawn in Section~\ref{s:end}.

%===============================================================================
\section{Formalism}
\label{s:eos}
\vspace{-6pt}

%-------------------------------------------------------------------------------
\subsection{Nuclear Equation of~State}

The nuclear EOS of the model is derived in the framework of the
Brueckner-Bethe-Goldstone theory,
which is based on a linked-cluster expansion of the energy per nucleon of
nuclear matter~\cite{1976Jeu,1999Book,2012Rep}.
The basic ingredient in this many-body approach is the in-medium Brueckner
reaction matrix $G$,
which~is the solution of the Bethe-Goldstone equation ($\hbar=c=1$)
\be
 G(E;\rho,x) = V + \sum_{1,2} V \frac
 {\left|1 2\right\rangle Q \left\langle 12\right|}
 { E - e_1 - e_2 } G(E;\rho,x) \:,
\label{e:g}
\ee
where $V$ is the bare nucleon-nucleon (NN) interaction,
$E$ is the starting energy,
and the multi-indices $1,2$ denote in general momentum, isospin, and~spin.
$x=\rho_p/\rho$ is the proton fraction, and~$\rho_p$ and $\rho$
are the proton and the total baryon density, respectively.
The propagation of intermediate baryon pairs is determined by the Pauli operator
$Q$ and the single-particle (s.p.) energy
\be
 e_1 = e(1;\rho,x) =
 \frac{k_1^2}{2m_1} + U_1 \:.
% \quad U_1(\rho,x) = {\rm Re} \sum_2
% n_2 \langle12|G(e_1+e_2;\rho,x)|12\rangle_a \:,
\label{e:e}
\ee
The BHF approximation for the s.p.~potential $U$
using the continuous choice is
\be
 U_1(\rho,x) = {\rm Re} \sum_{2<k_F^{(2)}}
 \langle 1 2| G(e_1+e_2;\rho,x) | 1 2 \rangle_a \:,
\label{e:u}
\ee
where the matrix element is antisymmetrized.
%as indicated by the ``$a$'' subscript.
Due to the occurrence of $U_1$ in Equation~(\ref{e:e}),
the~coupled system of Equations~(\ref{e:g})--(\ref{e:u})
must be solved in a self-consistent manner
for several Fermi momenta of the particles involved.
The corresponding BHF energy density is
\be
 \epsilon = \sum_{i=n,p} 2\sum_{k<k_F^{(i)}}
 \left( {k^2\over 2m_i} + {1\over 2}U_i(k) \right) \:.
\ee

It has been shown that the energy and the nuclear EOS
can be calculated with good accuracy
in the Brueckner two-hole-line approximation with the continuous choice for the
s.p.~potential,
since~the results in this scheme are quite close to the calculations
which include also the three-hole-line contribution~\cite{song1,song2,thl,thln}.
In this scheme,
the only input quantity needed is the bare NN interaction $V$
in the Bethe-Goldstone Equation~(\ref{e:g}).
In the present work, we use the Argonne $V_{18}$ potential~\cite{v18}
as the two-nucleon interaction,
supplemented by a consistent meson-exchange three-body force (TBF),
which allows to reproduce correctly the nuclear-matter
saturation point~\cite{1989Grange,2002Zuo,Li2008bp,zhl1}
and other properties of nuclear matter around saturation~\cite{jinbiao20}.

Further important ingredients in the cooling simulations
are the neutron and proton effective masses,
which we actually used in our previous simulations presented in
Reference~\cite{2016MNRAS}.
In the BHF approach,
the effective masses can be expressed self-consistently
in terms of the s.p.~energy $e(k)$ \cite{meff},
\be
 \frac{m^*(k)}{m} = \frac{k}{m} \left[ \frac{d e(k)}{dk} \right]^{-1} \:.
\ee
As found in~\cite{2016MNRAS},
their effect can be absorbed into a rescaling of the pairing gaps
that we also employ in this paper,
%is very small in comparison to other uncertainties regarding the cooling,
and therefore we simply use the bare nucleon mass here.
This is also convenient for comparison with other works that use bare~masses.

For completeness, we mention that
the BHF method provides the EOS for homogeneous nuclear matter,
$\rho > \rho_t \approx 0.08\fm3$.
For the low-density inhomogeneous crustal part
we adopt the well-known Negele-Vautherin EOS~\cite{NV}
for the inner crust in the medium-density regime
($0.001\fm3 < \rho < \rho_t$),
and the ones by Baym-Pethick-Sutherland~\cite{BPS} and
Feynman-Metropolis-Teller~\cite{fey} for the outer crust
($\rho < 0.001\fm3$).
The transition density $\rho_t$ is adjusted to provide a smooth transition
of pressure and energy density between both branches
of the betastable EOS~\cite{bhfls}.
The NS mass domain that we are interested in,
is hardly affected by the structure
of this low-density transition region and the crustal EOS:
The choice of the crust model can influence the predictions of
radius and related deformability to a small extent,
of the order of $1\%$ for
%the radius of a $M=1.4\ms$ NS,
the value of a 1.4-solar-mass NS,
$R_{1.4}$ \cite{bhfls,2014Bur_Cent,fortin},
which is negligible for our purpose.
Even neglecting the crust completely,
NS radius and deformability do not change dramatically~\cite{tsang19}.

\begin{figure}[H]%..............................................................
\vspace{-9mm}
\centerline{\includegraphics[scale=0.4]{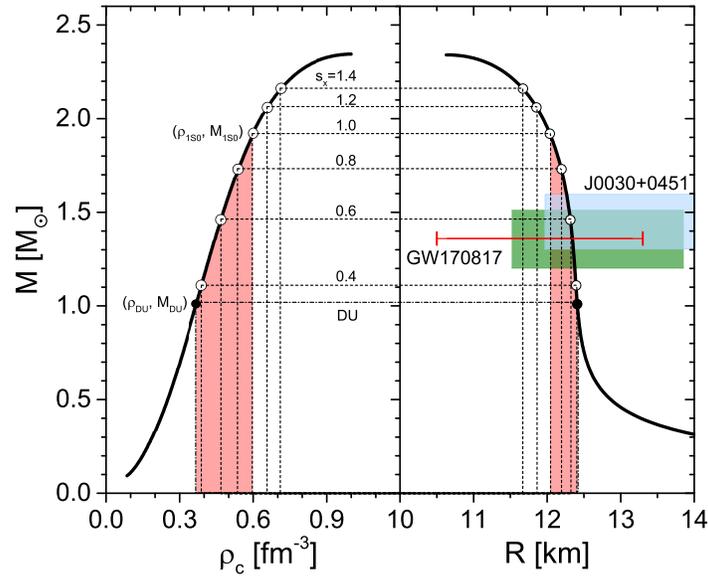}}
\vspace{-2mm}
\caption{
Neutron star gravitational mass vs central baryon density (left)
and radius (right) for the V18 EOS.
The density of the onset of DU cooling, $\rdu$,
and the one of the vanishing of the p1S0 BCS gap, $\r1s0$, are indicated.
The red-shaded region indicates blocking of the DU process.
The effect on $\r1s0$
of the pairing-range parameter $s_x$, Equation~(\ref{e:sxy}),
is also shown.
Some experimental constraints from NICER
(blue area~\cite{Miller_2019} or green area~\cite{Riley_2019})
and GW170817 (red bar) \cite{merger2}
are included.
}
\label{f:eos}
\end{figure}

In order to illustrate the bulk properties of the V18 EOS thus obtained,
Figure~\ref{f:eos} shows the resulting NS mass-radius and mass-central density
relations
obtained in the standard way by solving the TOV equations for betastable and
charge-neutral matter.
We remark that the value of the maximum mass $\mmax=2.34\ms$ of the V18 EOS
is larger than the current observational
lower limits~\cite{demorest2010,heavy2,fonseca16,cromartie}.
Regarding the radius,
we found in Reference~\cite{drago4,jinbiao} that for the V18 EOS
%the value of a 1.4-solar-mass NS,
$R_{1.4}=12.33$~km,
which fulfils the constraints derived from
the tidal deformability in the GW170817 merger event,
$R_{1.36}=11.9\pm1.4$~km~\cite{merger2},
see also similar compatible constraints on masses and radii
derived in References~\cite{%
Margalit17,  % MTOV < 2.17
Ruiz18,      % 2.16-2.28
Most18,      % 12.00 < R1.4 12.39 <13.45
Rezzolla18,  % 2.01-0.04+0.04 < MTOV < 2.16-0.15+0.17
Shibata19,   % MTOV < 2.3
Most20,      % MTOV > 2.08+0.04−0.04
Shao20}.     % MTOV = 2.13-0.08+0.09
The V18 EOS is also compatible with estimates of the mass and radius
of the isolated pulsar PSR J0030+0451
observed recently by NICER,
$M=1.44^{+0.15}_{-0.14}\ms$ and $R=13.02^{+1.24}_{-1.06}$~km~\cite{Miller_2019},
or~$M=1.36^{+0.15}_{-0.16}\ms$ and $R=12.71^{+1.14}_{-1.19}$~km~\cite{Riley_2019}.
The figure also shows
the density of the onset of direct Urca (DU) cooling, $\rdu$,
and the one of the vanishing of the p1S0 BCS gap, $\r1s0$,
to be introduced and discussed in the following,
along with the pairing parameter $s_x$.

%...................................................................

%-------------------------------------------------------------------------------
\subsection{Nuclear Cooling~Processes}
\label{s:cp}

One of the main cooling regulators,
besides the electromagnetic radiation from the surface,
is the neutrino emission from the NS core.
For sufficiently hot NSs the latter is in fact the main ingredient
of the NS cooling theory~\cite{2001rep,2006ARNPS,2006PaWe,2007LatPra,potrev15}.
Several different neutrino generation processes are possible inside NSs,
and their rates strongly depend on the NS EOS and composition and,
very important, on~the superfluid properties of the stellar matter, that is,
critical temperatures and gaps in the different pairing channels.
For instance, in~a non-superfluid NS,
the most powerful mechanism of neutrino emission is the direct Urca (DU) process,
which consists of a pair of charged weak-current reactions,
\be
 n \ra p + l + \bar{\nu}_l
\quad \textrm{and} \quad
 p + l \ra n + \nu_l \:,
\label{e:DU}
\ee
where $l$ is a lepton, electron or muon,
and $\nu_l$ is the corresponding neutrino.
Those reactions are allowed by energy and momentum conservation~\cite{1991LP}
only if $k_F^{(n)} < k_F^{(p)} + k_F^{(l)}$,
where $k_F^{(i)}$ is the Fermi momentum of the species $i$.
This implies that the proton fraction should be sufficiently high
for the DU process to take place,
and therefore the NS central density should be larger
than some threshold density.
Thus different EOSs predict different DU threshold~densities.

For some EOSs,
DU processes are forbidden in all NSs up to the most massive ones,
and then other reactions come into play.
In this case the basic neutrino emission mechanisms involve nucleon collisions,
the strongest one being the modified Urca (MU) process,
\be
 n + N \ra p + N + l \! + \bar{\nu}_l
\quad \textrm{and} \quad
 p + N + l \ra n + N + \nu_l \:,
\label{e:MU}
\ee
where $N$ is a spectator nucleon that ensures momentum conservation.
The nucleon-nucleon bremsstrahlung (BS) reactions,
\be
 N+N \ra N+N+ \nu+\bar{\nu} \:,
\ee
with $N$ a nucleon and $\nu$, $\bar{\nu}$ an (anti)neutrino of any flavor,
are also abundant in NS cores,
and their rate increases with the baryon density,
but they are orders of magnitude less powerful than the DU ones,
thus producing a slow cooling~\cite{2001rep}.

For the V18 EOS used in this work,
the DU process sets in at a proton fraction $x_p=0.135$
corresponding to a nucleon density $\rdu=0.37\fm3$
of beta-stable and charge-neutral matter,
and~an associated NS mass $\mdu=1.01\ms$,
as indicated in Figure~\ref{f:eos}.
Therefore practically all NSs can potentially cool very fast
and slow cooling has to be achieved by superfluid~suppression.

%===============================================================================
\subsection{Pairing Gaps and Critical~Temperatures}
\label{s:gap}

The neutrino cooling processes in NSs can be dramatically affected by
the neutron and proton superfluidity~\cite{2001rep,armen19},
and the knowledge of the pairing gaps $\Delta$ in the 1S0 and 3PF2 channels
in beta-stable matter is essential for a correct description
of the thermal evolution of a NS.
These superfluids are produced by the
$pp$ and $nn$ Cooper pairs formation
due to the attractive part of the NN potential,
and are characterized by a critical temperature $T_c \approx 0.567\Delta$
for isotropic gaps.
The occurrence of pairing leads to two relevant effects in NS cooling,
namely
(i) a strong reduction when $T < T_c$
of the emissivity of the neutrino processes the paired component is involved in,
with a corresponding reduction of the specific heat of that component;
(ii) onset of the ``Cooper pair breaking and formation'' (PBF) process
with associated neutrino-antineutrino pair emission.
This process starts when the temperature reaches $T_c$
of a given type of baryons,
becomes maximally efficient when $T \approx 0.8\,T_c$,
and then is exponentially suppressed for $T \ll T_c$ \cite{2001rep}.
%Of vital importance for any cooling simulation is the knowledge of the
%1S0 and 3PF2 pairing gaps for neutrons and protons in beta-stable matter,
%which on one hand block the DU and MU reactions,
%and on the other hand open new cooling channels by the PBF mechanism
%for stellar matter in the vicinity of the critical temperature~\cite{2001rep}.
As usual, we focus in this work on the most important proton 1S0 (p1S0)
and neutron 3PF2 (n3P2) pairing channels,
while the n1S0 (crust only) gap is much less relevant
for the cooling~\cite{2020MNRAS},
and the p3P2 gap is disregarded due to its uncertain properties
at extreme~densities.

In the simplest BCS approximation,
and detailing the more general case of pairing in the coupled 3PF2 channel,
the pairing gaps are computed by solving the (angle-averaged)
gap equation~\cite{bcsp1,bcsp2,bcsp3,bcsp4,bcsp5,bcsp6}
for the two-component $L=1,3$ gap function,
\be
  \left(\!\!\!\begin{array}{l} \de_1 \\ \de_3 \end{array}\!\!\!\right)\!(k) =
  - {1\over\pi} \int_0^{\infty}\!\! dk' {k'}^2 {1\over E(k')}
%\phantom{wwwwwwwww} && \nonumber\\ \times
  \left(\!\!\!\begin{array}{ll}
   V_{11}\!\! & \!\!V_{13} \\ V_{31}\!\! & \!\!V_{33}
  \end{array}\!\!\!\right)\!(k,k')
  \left(\!\!\!\begin{array}{l} \de_1 \\ \de_3 \end{array}\!\!\!\right)\!(k')
%&&
\label{e:gap}
\ee
with
\be
  E(k)^2 = [e(k)-\mu]^2 + \de_1(k)^2 + \de_3(k)^2 \:,
\ee
while fixing the (neutron or proton) density,
\be
  \rho = {k_F^3\over 3\pi^2}
   = 2 \sum_k {1\over 2} \left[ 1 - { e(k)-\mu \over E(k)} \right] \:.
\label{e:rho}
\ee
Here $e(k)=k^2\!/2m$ is the s.p.~energy,
$\mu \approx e(k_F)$ is the chemical potential
determined self-consistently from Equations~(\ref{e:gap})--(\ref{e:rho}),
and
\be
   V^{}_{LL'}(k,k') =
   \int_0^\infty \! dr\, r^2\, j_{L'}(k'r)\, V^{TS}_{LL'}(r)\, j_L(kr)
\label{e:v}
\ee
are the relevant potential matrix elements
($T=1$ and
$S=1$; $L,L'=1,3$ for the 3PF2 channel,
$S=0$; $L,L'=0$ for the 1S0 channel)
with the bare potential $V = V_{18}$.
The relation between (angle-averaged) pairing gap at zero temperature
$\de \equiv \sqrt{\de_1^2(k_F)+\de_3^2(k_F)}$
obtained in this way and the critical temperature of superfluidity is then
$T_c \approx 0.567\de$.
(If no angle average is performed, the~prefactor varies slightly,
see, for~example,~References~\cite{yakov99,2001rep}, % Table~1 ,potrev15},
but the angle-average procedure is usually an excellent
approximation~\cite{bcsp2,Papa17}).

However, in-medium effects might strongly modify these BCS results,
as both the s.p.~energy $e(k)$ and the interaction kernel $V$ itself
might include the effects of TBF and polarization corrections.
It turns out that in the p1S0 channel all these corrections lead
to a suppression of both magnitude and density domain of the BCS gap,
see, for~example, Figure~3 in Reference~\cite{ourgaps}
and Figure~3 in Reference~\cite{nspp07}
in the BHF context,
or Figure~7 in Reference~\cite{armen19}
and Figure~7 in Reference~\cite{nsbk01}
for a collection of different theoretical approaches,
while for the n3P2 channel both TBF and polarization effects on $V$
might be attractive and change the value of the gaps
even by orders of magnitude~\cite{ppol1,ppol2,ppol3}.
However, due to the high-density nature of this pairing,
all medium effects might be very strong and
there is still no reliable quantitative theoretical prediction for this~gap.
%MDPI: please confirm if the changes is OK
%AUTHORS: Yes, but removed the outer brackets (...) !?

\begin{figure}[H]%.............................................................
\centerline{\includegraphics[scale=0.4]{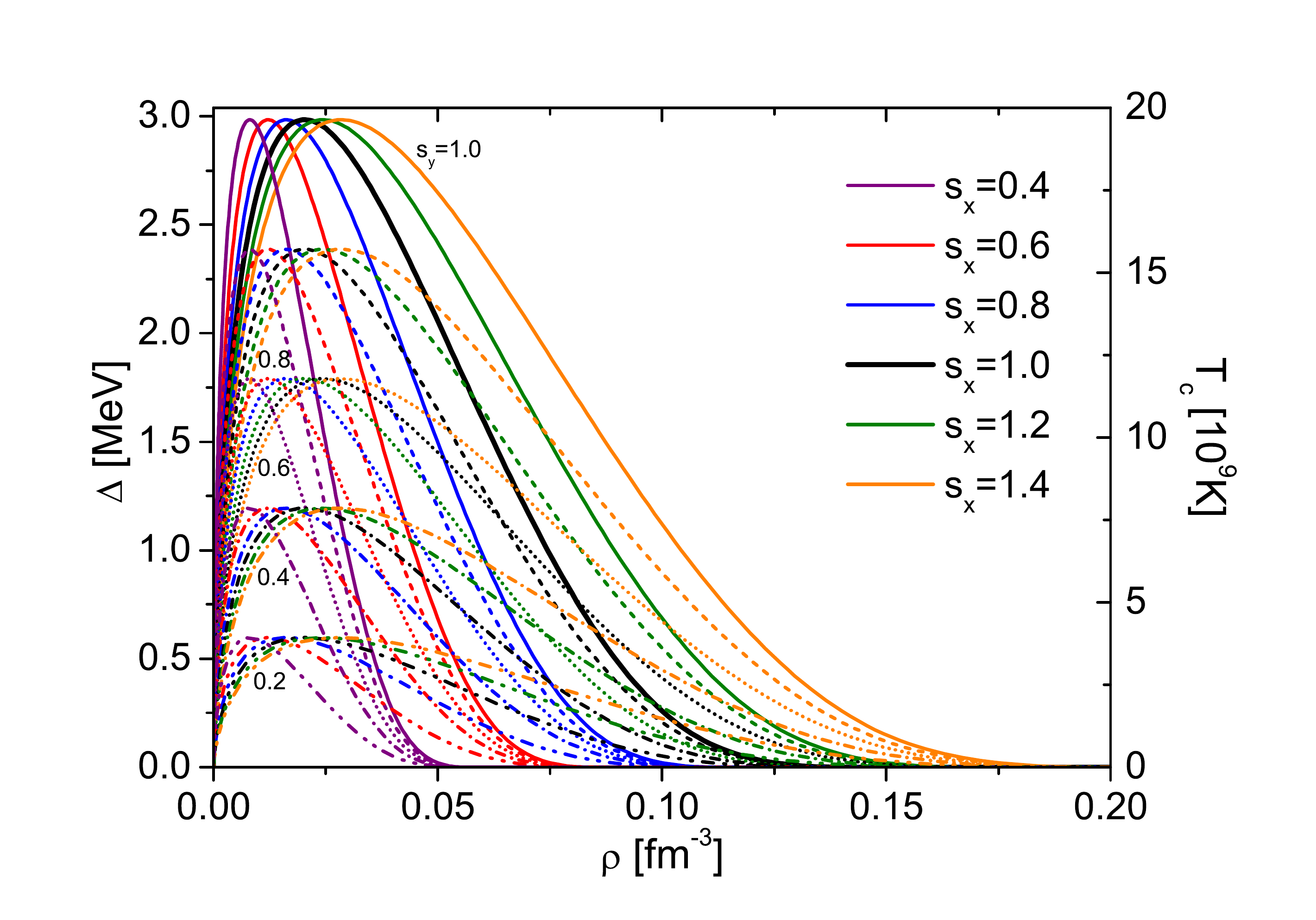}}
\vspace{-3mm}
\caption{
V18 BCS 1S0 pairing gap as a function of particle number density
with different scale factors $s_x$ and $s_y$ applied,
see Equation~(\ref{e:sxy}).
}
\label{f:gapp}
\end{figure}

We anticipate, however, that nonzero values of the n3P2 gap
cannot reproduce the current observational data in our cooling simulations,
so that we exclude it ad-hoc from now on
and concentrate our study on the possible p1S0 gap function $\Delta(\rho)$.
In this article we will not consider specific theoretical models
for the various medium effects,
but we use in the cooling simulations the density dependence of the
1S0 BCS pairing gap shown as solid black curve in Figure~\ref{f:gapp},
modified by simple global scaling factors $s_y$ and $s_x$
on either magnitude and extension of the gap, that is,
\be
 \Delta(\rho) \equiv s_y \Delta_\text{BCS}(\rho/s_x) \:.
\label{e:sxy}
\ee

The various trial gap functions are shown as colored curves
in Figure~\ref{f:gapp}
as a function of particle ($n$~or $p$) density
with different scale factors
$s_x=0.4,1.4(0.2)$ and $s_y=0.2,1.0(0.2)$ applied.
The~unscaled gap
($s_x=s_y=1$)
vanishes at $\rho\approx0.141\fm3$.
As just noted,
theoretical calculations point to a reduction of both magnitude and
density domain due to various in-medium effects
for this type of pairing,
which the scaling procedure attempts to mimic in a general way.
Thus scaling factors larger than one appear very unrealistic.
We include such choices only for the sake of a systematic investigation
of their effect on the cooling.
We now try to determine empirical optimal values of $s_x,s_y$
by comparing the theoretical results of our cooling simulations
with the currently known cooling~data.

\vspace{-12pt}
%..................................................................

%===============================================================================
\subsection{Cooling~Simulations}
\label{s:sim}

The NS cooling simulations are carried out using the
widely known code {\tt NSCool} \cite{Pageweb},
which~solves the general-relativistic equations
of energy balance and energy transport,
\bal
 \frac{1}{4\pi r^2e^{2\Phi}}
 \sqrt{1-\frac{2Gm}{r}} \frac{\partial}{\partial r}(e^{2\Phi}L_r) &=
 -Q_\nu - \frac{C_v}{e^\Phi} \frac{\partial T}{\partial t} \:,
\\
 \frac{L_r}{4\pi \kappa r^2} &=
 -\sqrt{1-\frac{2Gm}{r}} e^{-\Phi}\frac{\partial}{\partial r}(Te^\Phi) \:,
\eal
where $\Phi$ is the metric function,
$Q_\nu$, $C_v$, and~$\kappa$ are the neutrino emissivity,
the specific heat capacity, and~the thermal conductivity, respectively.
Local luminosity $L_r$ and temperature $T$ depend on
radial coordinate $r$ and time $t$.
In order to compute them,
the code adopts an implicit scheme (Henyey~scheme)
and solves the partial differential equations on a grid of spherical shells.
The shell number is 1842 in this work.
To facilitate the simulation,
the star is divided artificially at a outer boundary with the radius $r_b$
and density $\rho_b=10^{10}\gc3$.
At $\rho > \rho_b\ (r<r_b)$,
the matter is strongly degenerate
and thus the structure of the star is supposed to be unchanged with the time.
The envelope ($\rho<\rho_b$) includes the mass and composition change,
for instance, due to the accretion, and~is solved separately in the code.
Here, we use the envelope model obtained by Potekhin~\cite{Potekhin:1997mn}.
Each simulation starts with a constant initial temperature profile,
$\tilde{T}=Te^{\Phi}=10^{10}\,$K,
and ends when $\tilde{T}$ drops to $10^4\,$K.
Regarding the most important ingredient -- neutrino emissivity,
this code comprises all relevant cooling reactions:
nucleonic DU, MU, PBF, and~BS,
including modifications due to p1S0 and n3P2 pairing.
Moreover, various~processes in the crust are included,
such as the most important electron-nucleus bremsstrahlung,
plasmon decay, electron-ion bremsstrahlung, and~so~forth.

%==============================================================================
\section{Results}
\label{s:res}

\begin{figure}[H]%.............................................................
%\vspace{-10mm}
\centering
\includegraphics[scale=0.25]{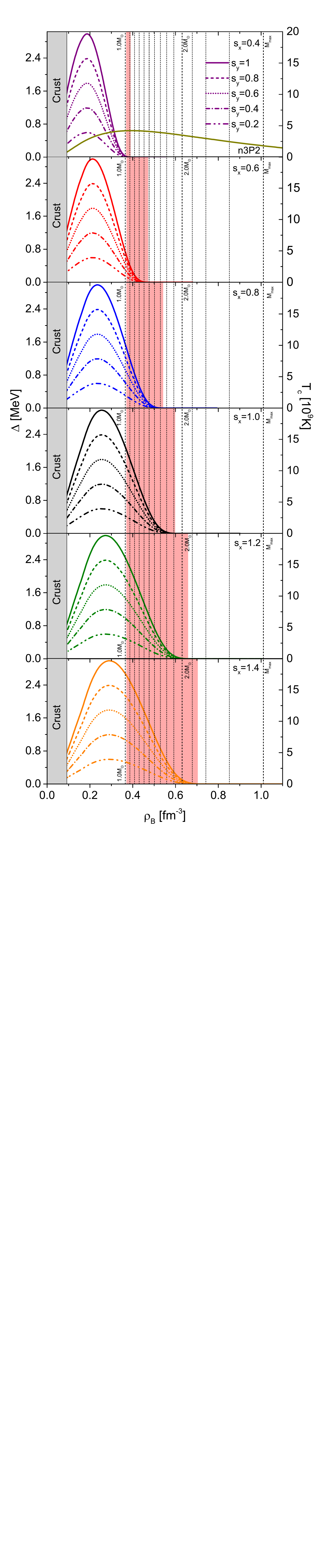}
\caption{V18 BCS p1S0 pairing gap in NS matter
as a function of baryon density
with different scale factors $s_x$ and $s_y$ applied.
The unscaled n3P2 gap is also shown for comparison
in the top panel.
Vertical~dotted lines indicate the central densities for
different NS masses $M/\!\ms=1.0,1.1,1.2,\ldots$.
The shaded red areas indicate the blocking of DU cooling.}
\label{f:gapb}
\end{figure}

Due to the density-dependent proton fraction,
the gap curves shown in Figure~\ref{f:gapp}
correspond to much wider domains of baryon density
for the p1S0 gap in NS matter
that are shown in Figure~\ref{f:gapb}.
That figure also indicates the central densities for several NS masses
(vertical lines)
and the mass ranges
(shaded red regions)
in which DU cooling is suppressed by the p1S0 gap.
Varying the $s_x$ parameter thus allows to regulate the NS mass domain
of DU blocking,
as was also indicated in Figure~\ref{f:eos}.
For the BCS case, $s_x=s_y=1$,
the p1S0 gap disappears at $\r1s0=0.60\fm3$,
corresponding to $\m1s0=1.92\ms$,
whereas the values for other scale factors $s_x$
are listed in Table~\ref{t:sx}.
From theoretical investigations~\cite{ourgaps,nspp07,nsbk01,armen19}
values of $s_x>1$ seem unrealistic,
but are nevertheless included in our systematic analysis.
For comparison also the unscaled n3P2 V18 BCS gap is shown in the figure.
It~extends over the entire density (mass) range and therefore
also would block DU cooling for all NSs.
However, the~competing n3P2 PBF process provides too strong cooling
for old objects,
in disagreement with some data,
as will be seen~later.

%..................................................................

%\squeezetable
\begin{table}[H]%..............................................................
\centering
\caption{
NS matter baryon densities $\r1s0$
for the vanishing of the p1S0 gap
and corresponding NS masses $\m1s0$
with that central density,
as a function of the scale parameter $s_x$.
}
%\medskip
\def\myc#1{\multicolumn{1}{c}{$#1$}}
\renewcommand{\arraystretch}{1.2}
\setlength{\tabcolsep}{5pt}
%\begin{ruledtabular}
\begin{tabular}{lccccccccccccc}
\toprule
\boldmath{$s_x$} &
\textbf{0.2} & \textbf{0.4} & \textbf{0.6} & \textbf{0.8} & \textbf{1.0} &
\textbf{1.2} & \textbf{1.4} & \textbf{1.6} &
\textbf{1.8} & \textbf{2.0} & \textbf{2.5} & \textbf{3.0} \\
\midrule
 $\r1s0$ $[\!\fm3]$ & 0.300& 0.388& 0.467& 0.536& 0.599 & 0.658& 0.713& 0.767
                    & 0.818& 0.869& 0.992& 1.114\\
 $\m1s0$ [$\!\ms$]  & 0.70 & 1.11 & 1.46 & 1.73  & 1.92 & 2.06 & 2.16 & 2.23
                    & 2.28 & 2.31 & 2.34 & 2.34 \\
\bottomrule
\end{tabular}
%\end{ruledtabular}
\label{t:sx}
\end{table}%...................................................................
%  V18   & 0.37 & 1.01  &   -   &   -   & 0.599 & 1.92   & 1.010 & 2.34\\

\begin{figure}[H]%.............................................................
\centering
\includegraphics[scale=0.8]{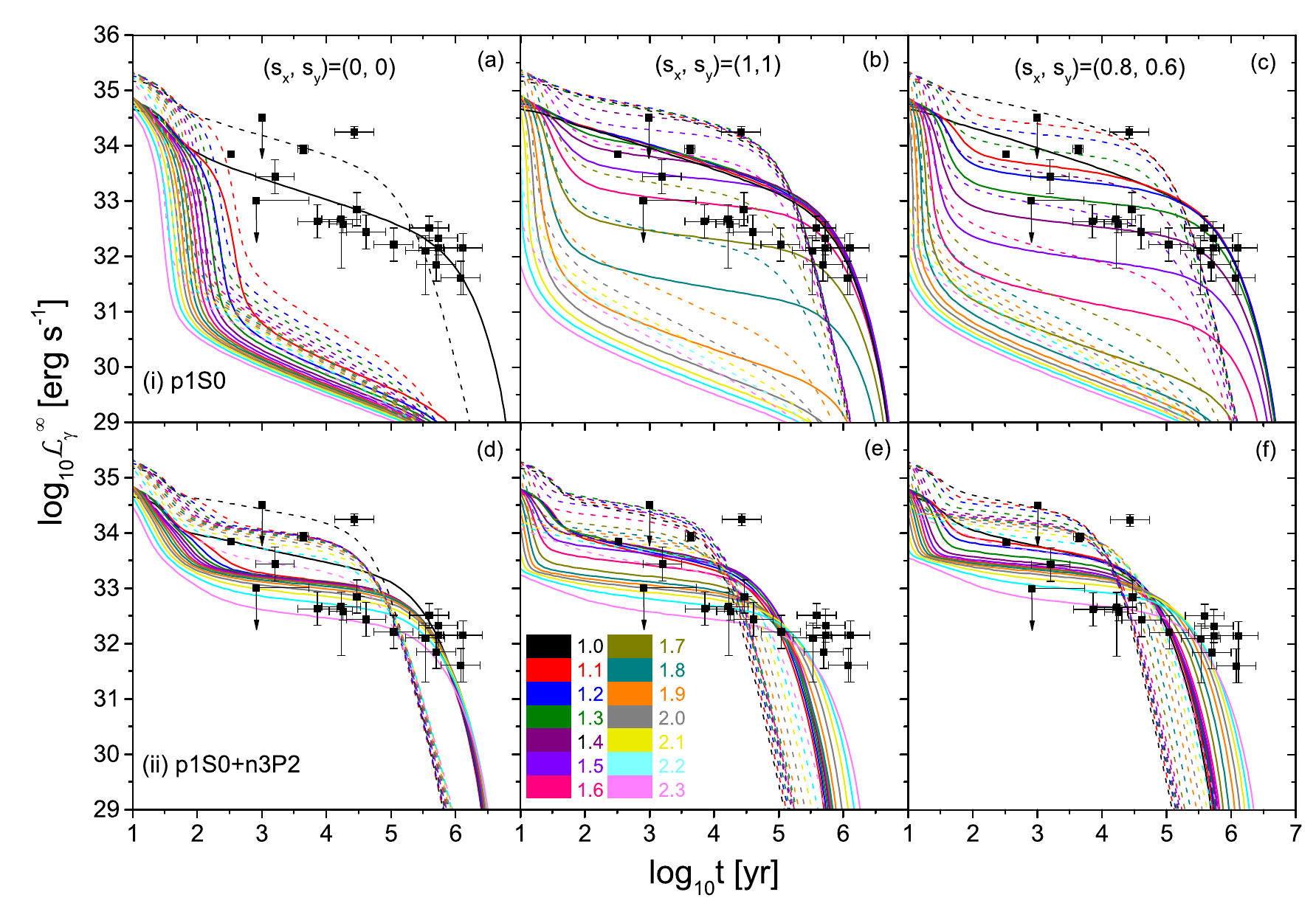}
\caption{
Cooling diagrams for p1S0 gap scaling factors
$(s_x,s_y)=(0,0);(1,1);(0.8,0.6)$
with (lower~row) and without (upper row) n3P2 pairing,
for different NS masses $M/\!\ms=1.0,1.1,\ldots,2.3$
(decreasing curves).
The solid curves are obtained with a Fe atmosphere
and the dashed curves
cover the same results obtained with a light-elements
($\eta=10^{-7}$) atmosphere.
The data points are from Reference~\cite{Beznogov1}.
See text for more details.
}
\label{f:cool}
%MDPI: please explain sub figure
%AUTHORS: Which sub figure?
\end{figure}

In Figure~\ref{f:cool} we show the cooling diagrams,
luminosity vs age,
for several sets $(s_x,s_y)$ of interest,
containing also the currently known data points with large
%(and partially only roughly
(estimated)
error bars, see Reference~\cite{Beznogov1}.
(We do not yet utilize the very recent update of these data~\cite{Potek20}
in this work.)
Theoretical results employing a Fe atmosphere (solid curves)
and those with a light-elements atmosphere (dashed curves)
are compared.
Starting with the absence of superfluidity (a), $s_x=s_y=0$, that is,
DU cooling active for all NSs with $M>1.01\ms$,
%apart from $M\approx\ms$,
one obtains clearly unrealistic results so that this scenario can be excluded.
Employing the original BCS values (b), $s_x=s_y=1$,
yields instead very reasonable results.
In this case only high-mass stars, $M>1.92\ms$, cool very rapidly,
whereas the cooling curves for lower masses
are smoothly distributed in the plot,
and not in apparent disaccordance with the data points.
Note that in this case all known cooling objects can be explained,
by assuming either a Fe or a light-elements atmosphere in specific cases,
even the very hot XMMU J1731-347,
($\log_{10}{t}\approx4.4$, $\log_{10}{L_\gamma^\infty}\approx34.2$),
which is indeed supposed to have a carbon atmosphere,
as discussed in References~\cite{Beznogov1,Beznogov2,Beznogov3,2018MNRAS}.
The~third set (c), $s_x=0.8,s_y=0.6$, is an optimal choice according
to the mass-distribution analysis presented in the following.
The difference to the BCS case is not very significant
and again all data can be~covered.

%..................................................................

While the upper-row panels (a,b,c) employ only the p1S0 gap,
the lower row (d,e,f) includes also the n3P2 BCS gap.
In this case, however, cooling is too fast due to the very efficient PBF
process of this channel,
see a detailed discussion in Reference~\cite{2020MNRAS},
and the luminosity of old ($\approx10^6$yr) NSs cannot be reproduced.
This problem also persists when allowing $s_y<1$ scale factors for
this channel~\cite{2016MNRAS,2018MNRAS}.
The~same conclusions were drawn by
References~\cite{Beznogov1,Beznogov2,Potek19,Potek20}
and other~authors.
%MDPI: please confirm if it should be Figure 4a--c and Figure 4 d--f
%AUTHORS: Yes.

We remark that the approach presented here
is able to describe perfectly with the same microphysics input
not only the cooling of isolated NSs discussed here,
but also the cooling of
reheated accreting NSs in X-ray transients in quiescence (XRTQ)
\cite{yako14,Beznogov1,Beznogov2},
as shown in Reference~\cite{2018MNRAS}.
We also remind the possible strong constraints on NS cooling imposed by the
speculated very rapid cooling of the
supernova remnant Cas~A~\cite{2009Nat,2010HeiHo,2013Elsha,2015HoPRC},
which we have studied in detail in Reference~\cite{2016MNRAS}.
As~the observational claims remain highly debated~\cite{casno1,casno2},
we do not consider this scenario in this~work.

The use of the V18 EOS together with the p1S0 BCS gap and no n3P2 pairing
appears thus to be consistent with the current set of cooling data.
However, without~information on the actual masses of the various data points,
it is currently impossible to perform a rigid check of the model by
comparing the theoretically predicted masses with the actual masses
of the data points in the cooling diagram.
In the absence of this possibility we dedicate this work
to another consistency check,
namely the theoretical derivation of the NS mass distribution
of the currently available data points
from their position among the theoretical cooling curves for different masses.
This mass distribution can then be compared to mass distributions of NSs
obtained in different, independent theoretical
ways~\cite{zhang11,ozel12,kizil13,anton16,alsing18,rocha19}.

Of course the validity of this analysis rests on two essential assumptions:
\begin{enumerate}[leftmargin=2.3em,labelsep=4mm]
\item[(a)] there is no bias on the masses (and thus luminosities)
in the current data set of isolated NSs, that~is,
bright and dim objects are supposed to be present with equal probability,
thus the detection of these sources is independent of their brightness;
\item[(b)] the mass distribution of isolated NSs in the cooling diagram is
not different from the mass distributions of NSs
in binary systems~\cite{zhang11,anton16,alsing18,rocha19,farrow19}
or all NSs in the Universe,
which were addressed in the other theoretical~studies.
\end{enumerate}

\begin{figure}[H]%...............................................................
\centering
\includegraphics[scale=1.5]{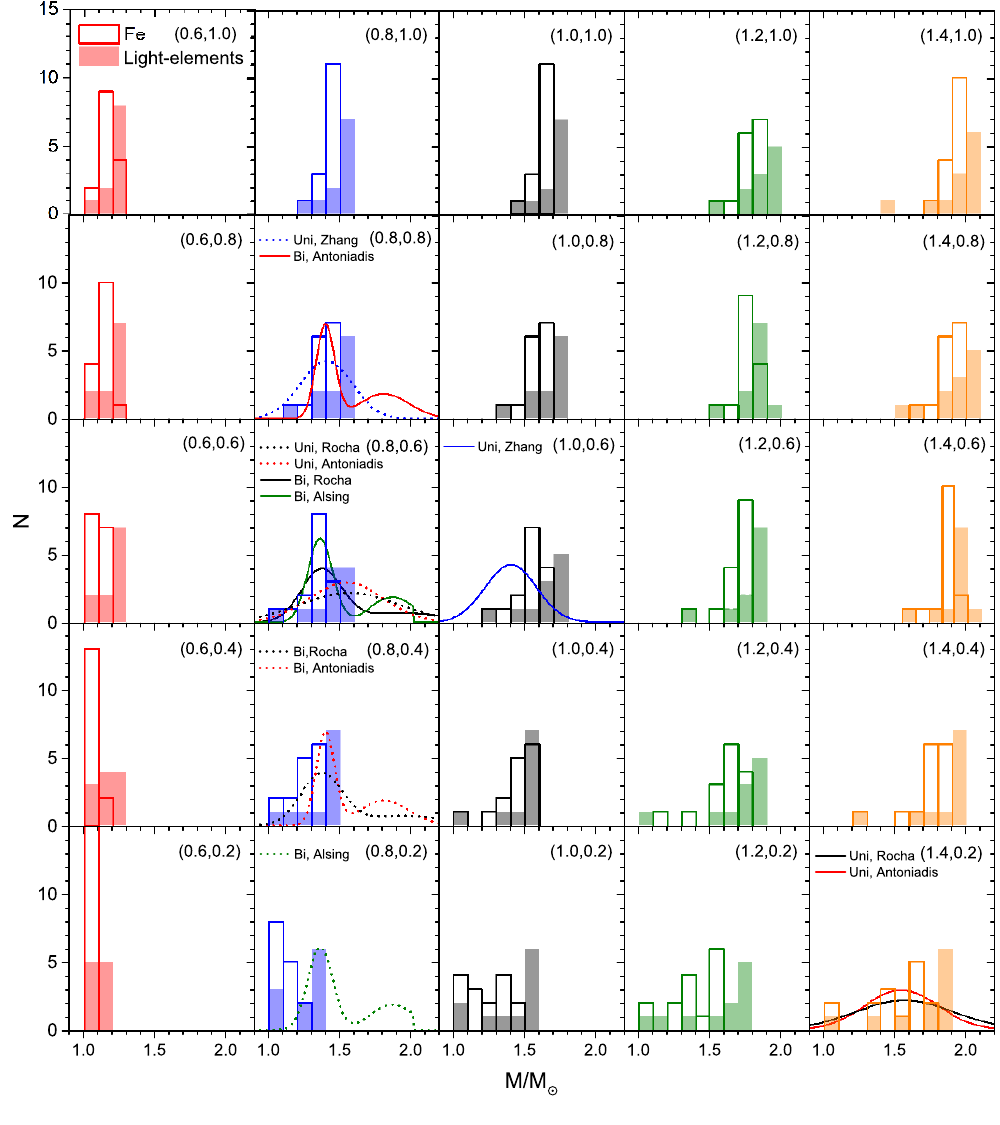}
\caption{
Deduced NS mass distributions for scaling factors
$(s_x=0.6,\ldots,1.4) \otimes (s_y=0.2,\ldots,1.0)$
and with Fe (solid lines)
or light-elements (shaded bars) atmosphere.
$N$ is the number of data points lying in a given mass interval
$\Delta M=0.1\ms$
in the proper ($s_x,s_y$) cooling diagram.
Several panels show the best-fit theoretical
results~\cite{zhang11,anton16,alsing18,rocha19}
of Figure~\ref{f:theo} superimposed,
using solid (dotted) lines for the Fe (light-elements) results.
}
\label{f:mass}
\end{figure}

Both of these assumptions are highly unlikely to be fulfilled,
but are currently impossible to further scrutinize quantitatively.
We therefore proceed with our derivation of the mass distribution of
the current data as outlined above.
The masses of the 19 cooling data points are assumed to be those predicted
theoretically by their position in the cooling diagrams
among the theoretical curves,
see, for~example, Figure~\ref{f:cool}
(neglecting any error bars at this stage of investigation),
and Figure~\ref{f:mass} shows the resulting mass histograms for different
choices of the parameters $s_x,s_y$,
and assuming either a common Fe or a light-elements atmosphere
for all data points.
This is clearly unrealistic,
and in fact in the first case only 15 and in the second case only 11
[apart from 10 for (0.6,0.2) and 12 for (1.4,0.8) and (1.4,1.0)]
sources can be fitted, see Figure~\ref{f:cool},
while a fit of all data would require a suitable choice of atmosphere
for each object.
One observes that increasing $s_x$ or to a lesser degree $s_y$
shifts the centroid of the derived mass distribution to higher values,
since the cooling curves like in Figure~\ref{f:cool}
shift upwards due to the increased suppression of the DU~process.

\begin{figure}[H]%.............................................................
\vspace{-7mm}
\centerline{\includegraphics[scale=0.8]{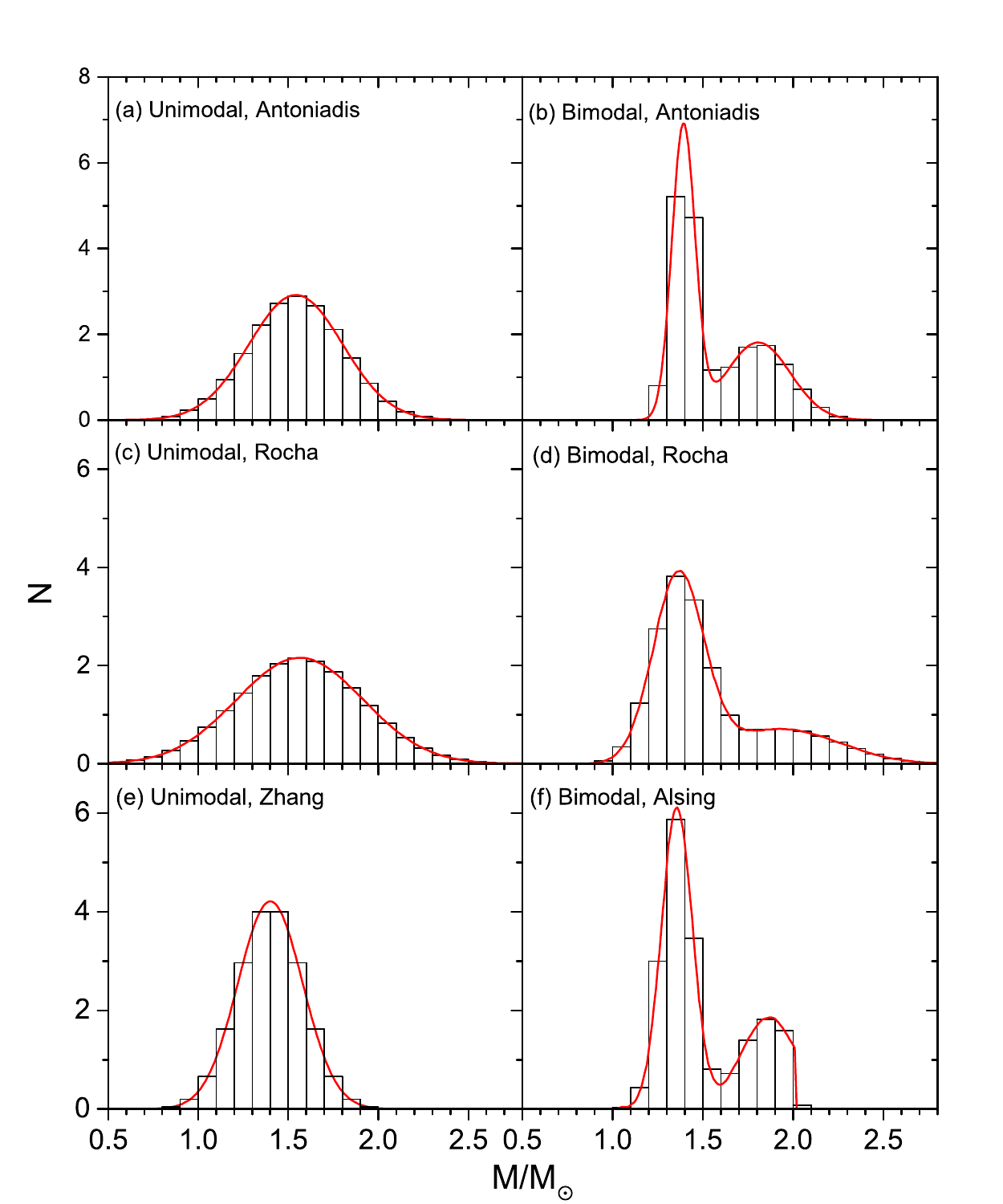}}
\vspace{-1mm}
\caption{
Theoretical NS mass distributions by several
authors~\cite{zhang11,anton16,alsing18,rocha19}
(red curves)
binned and normalized in the same way as Figure~\ref{f:mass}
(black histograms).
}
\label{f:theo}
\end{figure}

The different results in Figure~\ref{f:mass} can now be compared with
other theoretical studies of the NS mass distribution.
There is a long-lasting discussion whether this distribution is unimodal
or bimodal due to different classes of NS evolution
histories~\cite{zhang11,ozel12,kizil13,anton16,alsing18,rocha19},
and in Figure~\ref{f:theo} we show a compilation of some recent
theoretical studies for the NS mass distribution.
We provide a binning consistent with Figure~\ref{f:mass}
in order to compare directly with our results.
For a quantitative comparison we simply compute the rms deviations between
the histograms in Figure~\ref{f:theo} and those in Figure~\ref{f:mass}.
The results are given in Table~\ref{t:rms} for the various combinations,
where we also indicate the optimal values $s_x,s_y$
for each theoretical mass distribution
and the two atmosphere models separately.
Of course the use of a unique atmosphere model for the whole data set
is unrealistic,
but currently impossible to overcome without further constraints on the~data.

%..................................................................

Nevertheless some qualitative conclusions can be drawn.
Whereas for a Fe atmosphere
a good agreement with some unimodal distributions
seems to require fairly large (and unphysical) values of $s_x\gtrsim 1$
in order to shift the median to sufficiently large mass
$\bar{M}\approx1.5\ms$,
%the confrontation with bimodal distributions provides
most other cases indicate
clearly preferred values of $s_x\approx0.8$ and $s_y\approx0.6\dots0.8$,
which would also be more consistent with microscopic investigations
of the 1S0 pairing gap, as~discussed before.
The quality of agreement is only slightly better for unimodal distributions,
but for too large $s_x$.
For a light-elements atmosphere,
{\em all} theoretical models single out $s_x=0.8$ as preferred value,
while $s_y$ is not well~determined.

The ($s_x,s_y$) configurations which fit best a given theoretical model,
have that theoretical curve superimposed in Figure~\ref{f:mass},
and four of the models
(two unimodal, two bimodal)
single out (0.8,0.6) as `best' parameter set.
Those values (0.8,0.6) were also used for some cooling diagrams
shown in Figure~\ref{f:cool}.
We conclude that in nearly all cases
the comparison between cooling diagrams and population models
indicates a reduction of the pairing range to $s_x\approx0.8$
and also a reduction of the magnitude $s_y\approx0.6$,
which is however less well determined and more model dependent.
Both features are also supported by theoretical predictions for the 1S0 gap.
The current limitations of theoretical method and available data
do not allow to single out a preferred theoretical population model,
though.

\begin{table}[H]%..............................................................
\centering
\caption{
%RMS %please define
Root-mean-square
variances between deduced NS mass distributions
with different p1S0 gap scale factors $s_x,s_y$
for Fe or light-elements atmospheres,
shown in Figure~\ref{f:mass},
and different theoretical models~\cite{zhang11,anton16,alsing18,rocha19},
shown in Figure~\ref{f:theo}.
Preferred values are indicated in boldface.
}
%\medskip
\def\myc#1{\multicolumn{1}{c}{$#1$}}
\renewcommand{\arraystretch}{1.2}
\setlength{\tabcolsep}{5pt}
%\begin{ruledtabular}
\begin{tabular}{lccccccccccc}
\toprule
 & \textbf{Atmosphere} & \multicolumn{5}{c}{\textbf{Fe}}
 & \multicolumn{5}{c}{\textbf{Light Elements}} \\
\midrule
 & $s_y \backslash s_x$
       & 0.6  & 0.8  & 1.0  & 1.2  & 1.4  & 0.6  & 0.8  & 1.0  & 1.2  & 1.4  \\
\midrule
 \multirow{5}{3.5cm}{Unimodal~Antoniadis~\cite{anton16}}
 & 1.0 & 2.46 & 2.28 & 2.22 & 1.96 & 2.61 & 2.09 & 1.39 & 1.54 & 1.60 & 1.36 \\
 & 0.8 & 2.69 & 1.77 & 1.56 & 2.07 & 2.20 & 1.95 & 1.23 & 1.33 & 1.79 & 1.53 \\
 & 0.6 & 2.69 & 1.78 & 1.28 & 1.42 & 2.33 & 1.95&\bf1.07& 1.17 & 1.76 & 1.87 \\
 & 0.4 & 3.29 & 1.83 & 1.31 & 1.28 & 1.78 & 1.78 & 1.55 & 1.48 & 1.43 & 1.90 \\
 & 0.2 & 3.72 & 2.45 & 1.55 & 1.37&\bf1.00& 2.03 & 1.59 & 1.34 & 1.22 & 1.57 \\
\midrule
 \multirow{5}{3.5cm}{Unimodal Rocha~\cite{rocha19}}
 & 1.0 & 1.92 & 1.99 & 1.94 & 1.57 & 2.03 & 1.64 & 1.23 & 1.26 & 1.17 & 1.38 \\
 & 0.8 & 2.12 & 1.54 & 1.43 & 1.69 & 1.68 & 1.50 & 1.08 & 1.10 & 1.36 & 1.23 \\
 & 0.6 & 2.11 & 1.49 & 1.22 & 1.65 & 1.85 & 1.48&\bf0.86& 1.43 & 1.35 & 1.42 \\
 & 0.4 & 2.63 & 1.45 & 1.22 & 1.14 & 1.41 & 1.54 & 1.29 & 1.27 & 1.06 & 1.41 \\
 & 0.2 & 3.00 & 1.90 & 1.20 & 1.20&\bf0.90& 1.18 & 1.34 & 1.13 & 0.95 & 1.19 \\
\midrule
 \multirow{5}{3.5cm}{Unimodal Zhang~\cite{zhang11}}
 & 1.0 & 2.37 & 2.48 & 2.48 & 2.61 & 2.80 & 1.97 & 1.48 & 2.12 & 2.17 & 2.28 \\
 & 0.8 & 2.68 & 1.61 & 1.95 & 2.71 & 1.95 & 1.86 &\bf1.29&1.92 & 2.34 & 2.08 \\
 & 0.6 & 2.74 & 1.76&\bf1.10& 2.54 & 2.42 & 1.86 & 1.33 & 1.77 & 2.29 & 2.28 \\
 & 0.4 & 3.38 & 1.45 & 1.14 & 1.90 & 2.15 & 1.77 & 1.38 & 1.65 & 2.06 & 2.33 \\
 & 0.2 & 3.81 & 2.47 & 1.29 & 1.22 & 1.40 & 2.15 & 1.51 & 1.47 & 1.78 & 2.09 \\
\midrule
 \multirow{5}{3.5cm}{Bimodal~Antoniadis~\cite{anton16}}
 & 1.0 & 3.22 & 1.92 & 3.15 & 2.56 & 3.00 & 3.32 & 2.55 & 2.68 & 2.52 & 2.15 \\
 & 0.8 & 3.43&\bf1.09& 2.54 & 2.72 & 2.62 & 3.18 & 2.23 & 2.49 & 2.75 & 2.36 \\
 & 0.6 & 3.40 & 1.30 & 2.24 & 2.19 & 2.83 & 3.18 & 1.87 & 2.39 & 2.61 & 2.69 \\
 & 0.4 & 3.95 & 1.99 & 1.72 & 2.29 & 2.43 & 3.00&\bf1.84& 2.70 & 2.49 & 2.83 \\
 & 0.2 & 4.37 & 3.17 & 1.76 & 1.90 & 1.58 & 3.24 & 2.05 & 2.55 & 2.37 & 2.45 \\
\midrule
 \multirow{5}{3.5cm}{Bimodal Rocha~\cite{rocha19}}
 & 1.0 & 2.11 & 1.80 & 2.46 & 2.19 & 2.51 & 1.67 & 1.40 & 1.82 & 1.70 & 1.77 \\
 & 0.8 & 2.40 & 1.14 & 1.88 & 2.31 & 2.21 & 1.57 & 1.22 & 1.64 & 1.90 & 1.69 \\
 & 0.6 & 2.45&\bf1.09& 1.53 & 2.23 & 2.39 & 1.57 & 1.25 & 1.51 & 1.85 & 1.88 \\
 & 0.4 & 3.03 & 1.20 & 1.19 & 1.73 & 2.01 & 1.50 &\bf1.21&1.54 & 1.67 & 1.85 \\
 & 0.2 & 3.42 & 2.20 & 1.09 & 1.17 & 1.28 & 1.86 & 1.21 & 1.36 & 1.50 & 1.68 \\
\midrule
 \multirow{5}{3.5cm}{Bimodal Alsing~\cite{alsing18}}
 & 1.0 & 3.49 & 2.66 & 3.91 & 3.11 & 3.47 & 2.76 & 2.64 & 2.84 & 2.52 & 2.21 \\
 & 0.8 & 3.89 & 1.53 & 3.21 & 3.32 & 3.07 & 2.66 & 2.37 & 2.61 & 2.76 & 2.41 \\
 & 0.6 & 3.94 &\bf1.20&2.81 & 2.74 & 3.34 & 2.66 & 2.08 & 2.50 & 2.62 & 2.64 \\
 & 0.4 & 4.62 & 1.69 & 2.29 & 2.87 & 2.89 & 2.66 & 2.14 & 2.80 & 2.56 & 2.72 \\
 & 0.2 & 5.12 & 3.58 & 1.86 & 2.13 & 2.19 & 3.16&\bf1.68& 2.58 & 2.46 & 2.46 \\
\bottomrule
\end{tabular}
\label{t:rms}
\end{table}%...................................................................

%..................................................................

%===============================================================================
\section{Conclusions}
\label{s:end}

In this model study we have investigated the important effect of
neutron and proton pairing gaps on NS cooling,
which suppress the dominant DU process
and open the competing PBF cooling reactions.
For the betastable matter and resulting NS structure
we used a fixed microscopically-derived BHF EOS
that fulfils all current phenomenological constraints.
This EOS features DU cooling for practically all NSs,
which has to be (partially) quenched by~superfluidity.

However, neutron 3P2 pairing leads to too rapid cooling of old NSs
by the PBF process in contrast to several known objects
and has to be excluded within this model,
while proton 1S0 pairing together with an adjustment of the NS atmosphere
is able to provide a satisfactory description of all current data.
We then computed the derived NS mass distribution
of the present set of data points
from the cooling diagrams
as a functional of the proton 1S0 gap,
by scaling both the magnitude and the density domain of the BCS gap.
In this way optimal scale factors
(pointing to a reduction of both magnitude and density domain)
and a optimal gap function could be obtained
by reproducing a given theoretical mass distribution.
The analysis yields a reasonable agreement with current theoretical
bimodal mass distributions,
due to the fact that the dominant peak is located at lower mass
than in the unimodal distributions.
However, unique atmosphere models for all sources
had to be assumed in this~setup.

Thus, at~the current status of observational data from isolated NSs
(few data points with large errors,
not providing essential information on the NS masses and atmospheres),
as well as an equally uncertain theoretical mass distribution of isolated NSs,
this work can only be a `proof of concept' study within our approach
(following the works of~\cite{Beznogov1,Beznogov2,Beznogov3})
that demonstrates the potential that high-quality data would have to
constrain the combination of nuclear bulk EOS and superfluid gaps
in the future.
This would complement the constraints obtained from other sources
(on masses, radii, deformability, glitches, etc.)
and ultimately allow the identification of the `unique' NS EOS
and superfluid properties of nuclear matter.
With always more precise satellite observations,
this goal might be achievable in the~future.

%===============================================================================
\vspace{6pt}
\authorcontributions{All authors contributed to this work.
All authors have read and agreed to the published version of the manuscript.}
%please state individual contribution
%AUTHORS: It is fine like this.

%\supplementary{The following are available online at \linksupplementary{s1},
%Figure S1: title, Table S1: title, Video S1: title.}

\funding{% https://search.crossref.org/funding
This work was partially funded by
``PHAROS'' COST Action CA16214
and the China Scholarship Council, No.~201706410092.
}

\acknowledgments{
We acknowledge helpful communication with D.~Page.
}

\conflictsofinterest{The authors declare no conflicts of~interest.}

%\appendixtitles{no}
%\appendix
%\section{}

%===============================================================================
%http://www.issn.org/services/online-services/access-to-the-ltwa/
\newcommand\mathplus{+}
\newcommand\mdash{--}
\newcommand{\araa}{Annu. Rev. Astron.~Astrophys.\ }
\newcommand{\aap}{Astron.~Astrophys.\ }
\newcommand{\apjl}{Astrophys.~J.~Lett.\ }
\newcommand{\apj}{Astrophys.~J.\ }
\newcommand{\epja}{EPJA\ }
\newcommand{\mnras}{Mon. Not. R. Astron.~Soc.\ }
\newcommand{\nat}{Nature\ }
\newcommand{\nphysa}{Nucl.~Phys.~A\ }
\newcommand{\physrep}{Phys.~Rep.\ }
\newcommand{\plb}{Phys.~Lett.~B\ }
\newcommand{\prc}{Phys.~Rev.~C\ }
\newcommand{\prd}{Phys.~Rev.~D\ }
\newcommand{\prl}{Phys.~Rev.~Lett.\ }
\newcommand{\ssr}{Space~Sci.~Rev.\ }

\reftitle{References}

\end{document}